\documentclass{nature_1sp}[12pt]
\usepackage[dvipsnames]{xcolor}
\usepackage{afterpage}
\usepackage{color}
\usepackage{amssymb,times}
\usepackage{amsmath}
\usepackage[english]{babel}
\usepackage{subfigure}
\usepackage{ifpdf}
\usepackage[rightcaption]{sidecap}
\usepackage{mathtools,array}
\usepackage{media9}
\usepackage[colorlinks = true,
            linkcolor = black,
            urlcolor  = black,
            citecolor = black,
            anchorcolor = black]{hyperref}
            
\usepackage{soul}
\usepackage{tcolorbox}

\newcount\longrefs
\def\aap{\ifnum\longrefs=1 {Astron.\ Astrophys.}\else 
                           {A\hbox{\rm \&}A}\fi}
\def\aapl{\ifnum\longrefs=1 {Astron.\ Astrophys.\ Lett.}\else 
                           {A\hbox{\rm \&}A}\fi}
\def\aapr{\ifnum\longrefs=1 {Astron.\ Astrophys.\ Rev.}\else 
                            {A\hbox{\rm \&}AR}\fi}
\def\aaps{\ifnum\longrefs=1 {Astron.\ Astrophys.\ Suppl.}\else 
                            {A\hbox{\rm \&}AS}\fi}
\def\aj{\ifnum\longrefs=1 {Astron.\ J.}\else 
                          {AJ}\fi} 
\def\ao{\ifnum\longrefs=1 {Applied Optics}\else 
                           {Appl.\ Opt.}\fi} 
\def\aspcs{\ifnum\longrefs=1 {Astron.\ Soc.\ Pacific Conf. Series}\else 
                           {ASP Conf.\ Ser.}\fi} 
\def\apj{\ifnum\longrefs=1 {Astrophys.\ J.}\else 
                           {ApJ}\fi} 
\def\apjl{\ifnum\longrefs=1 {Astrophys.\ J.\ Lett.}\else 
                            {ApJ}\fi} 
\def\aplett{\ifnum\longrefs=1 {Astrophys.\ J.\ Lett.}\else 
                            {ApJ}\fi} 
\def\apjs{\ifnum\longrefs=1 {Astrophys.\ J.\ Suppl.}\else 
                            {ApJS}\fi}
\def\apss{\ifnum\longrefs=1 {Astrophys.\ and Space Science}\else 
                            {Ap\hbox{\rm \&}SS}\fi}
\def\araa{\ifnum\longrefs=1 {Ann.\ Rev.\ Astron.\ Astrophys.}\else 
                            {ARA\hbox{\rm \&}A}\fi}
\def\azh{\ifnum\longrefs=1 {Astronomicheskii Zhurnal}\else 
                            {Astron.\ Zhur.}\fi}
\def\baas{\ifnum\longrefs=1 {Bull.\ Am.\ Astron.\ Soc.}\else 
                            {BAAS}\fi}
\def\bain{\ifnum\longrefs=1 {Bull.\ Astronom.\ Institutes Netherlands}\else
                            {Bull.\ Astr.\ Inst.\ Neth.}\fi}
\def\gca{\ifnum\longrefs=1 {Geochim.\ Cosmochim.\ Acta}\else 
                           {Geochim.\ Cosmochim.\ Acta}\fi}
\def\grl{\ifnum\longrefs=1 {Geophys.\ Res.\ Lett.}\else 
                           {Geoph.\ Res.\ Lett.}\fi}
\def\iaucirc{\ifnum\longrefs=1 {IAU Circulars}\else 
                          {IAU Circ.}\fi}
\def\ip{\ifnum\longrefs=1 {in press}\else 
                          {in press}\fi}
\def\jchemp{\ifnum\longrefs=1 {J.\ Chem.\ Phys.}\else 
                           {J.\ Chem.\ Phys.}\fi}  
\def\jcp{\ifnum\longrefs=1 {J.\ Chem.\ Phys.}\else 
                           {J.\ Chem.\ Phys.}\fi}  
\def\jgr{\ifnum\longrefs=1 {J.\ Geophys.\ Res.}\else 
                           {J.\ Geophys.\ Res.}\fi}  
\def\jmolspec{\ifnum\longrefs=1 {J.\ Mol.\ Spectrosc.}\else 
                           {J.\ Mol.\ Spectrosc.}\fi}  
\def\jqsrt{\ifnum\longrefs=1 {J.\ Quant.\ Spectrosc.\ Radiat.\ Transfer}\else 
                           {J.\ Quant.\ Spectrosc.\ Radiat.\ Transfer}\fi}  
\def\jrasc{\ifnum\longrefs=1 {J.\ Royal Astron.\ Soc.\ Canada}\else 
                           {JRAS Can.}\fi}  
\def\mnras{\ifnum\longrefs=1 {Mon.\ Not.\ Roy.\ Astron.\ Soc.}\else 
                             {MNRAS}\fi} 
\def\nat{\ifnum\longrefs=1 {Nature}\else 
                           {Nat}\fi}
\def\pasj{\ifnum\longrefs=1 {Pub.\ Astron.\ Soc.\ Japan}\else 
                            {PASJ}\fi} 
\def\pasp{\ifnum\longrefs=1 {Pub.\ Astron.\ Soc.\ Pacific}\else 
                            {PASP}\fi} 
\def\physscr{\ifnum\longrefs=1 {Physica Scripta}\else 
                            {Phys.\ Scrip.}\fi} 
\def\planss{\ifnum\longrefs=1 {Planetary \& Space Science}\else 
                            {Plan. \& Space Sci.}\fi} 
\def\procspie{\ifnum\longrefs=1 {Proc.\ SPIE}\else 
                            {Proc.\ SPIE}\fi} 
\def\qjras{\ifnum\longrefs=1 {Quarterly J.\ Royal Astron.\ Soc.}\else 
                            {QJRAS}\fi} 
\def\sa{\ifnum\longrefs=1 {Soviet Astron..}\else 
                               {Sov.\ Astron.}\fi}
\def\skytel{\ifnum\longrefs=1 {Sky \& Telescope}\else 
                            {Sky \& Tel.}\fi} 
\def\solphys{\ifnum\longrefs=1 {Solar Phys.}\else 
                               {Solar Phys.}\fi}
\def\ssr{\ifnum\longrefs=1 {Space Science Rev.}\else 
                               {Space\ Sci.\ Rev.}\fi}

\def\Msun{\hbox{M$_{\odot}$}}               
\def\Rstar{\hbox{R$_{\star}$}}              
\def\Mdot{\hbox{$\dot{M}$}}               
\def\arcsec{\hbox{$^{\prime\prime}$}}

\def\oh{\hbox{OH\,26.5+0.6}}
\def\OH30{\hbox{OH\,30.1$-$0.7}}
\def\farcs{\mbox{$.\!\!^{\prime\prime}$}}%

\title{Reduced maximum mass-loss rate of OH/IR stars due to overlooked binary interaction}

\author{L.\ Decin$^{1,2,\star}$,
	W.\ Homan$^{1}$,
        T.\ Danilovich$^{1}$,
        A.\ de Koter$^{1,3}$,
 	D.~Engels$^{4}$,
 	L.~B.~F.~M.\ Waters$^{3,5}$,
 	S.\ Muller$^{6}$,
        C.\ Gielen$^{1}$,
 	D.~A.\ Garc\'{\i}a-Hern\'{a}ndez$^{7,8}$,
 	R.~J.\ Stancliffe$^{9,10}$,
  	M.\ Van de Sande$^{1}$,
  	G.\ Molenberghs$^{11,12}$,
 	F.\ Kerschbaum$^{13}$,
 	A.~A.\ Zijlstra$^{14,15}$
 	\&
 	I.\ El Mellah$^{16}$
 }

\tcbset{width=0.9\textwidth,boxrule=0pt,colback=lightgray,arc=0pt,auto outer arc,left=0pt,right=0pt,boxsep=5pt}

\begin{document}
\newcommand{\red}{\textcolor[rgb]{1,0,0}}
\newcommand{\blue}{\textcolor[rgb]{0,0,1}}
\newcommand{\BlueGreen}{\textcolor[rgb]{0,1,1}}
\newcommand{\orange}{\textcolor[rgb]{1,0.5,0}}

\maketitle

\small{
\begin{affiliations}
	\item Instituut voor Sterrenkunde, KU Leuven, Celestijnenlaan 200D, 3001 Leuven, Belgium\\ $^\star$e-mail: leen.decin@kuleuven.be
 		\item
	University of Leeds, School of Chemistry, Leeds LS2 9JT, United Kingdom 
	\item
	Astronomical Institute Anton Pannekoek, University of Amsterdam, Science Park 904, PO Box 94249, 1090 GE, Amsterdam, The Netherlands
	   \item
	   Hamburger Sternwarte, Gojenbergsweg 112, D-21029 Hamburg, Germany
	   \item
	   SRON Netherlands Institute for Space Research, Sorbonnelaan 2, 3584 CA Utrecht, The Netherlands
  	\item
	   Chalmers University of Technology, Department of Space, Earth and Environmnet, Onsala Space Observatory, S-439 92 Onsala, Sweden
	   \item
	   Instituto de Astrof\'{\i}sica de Canarias, E-38205 La Laguna, Tenerife, Spain
	   \item
	   Departamento de Astrof\'{\i}sica, Universidad de La Laguna (ULL), E-38206 La Laguna, Tenerife, Spain
	   \item
	   E.\ A.\ Milne Centre for Astrophysics, Department of Physics \& Mathematics, University of Hull, HU6 7RX, UK
	   \item
	   School of Physics and Astronomy, University of Birmingham, Birmingham B15 2TT, UK
       \item
       I-BioStat, Universiteit Hasselt, Martelarenlaan 42, 3500, Hasselt, Belgium
       \item I-BioStat, KU Leuven, Kapucijnenvoer 35, 3000, Leuven, Belgium	   
	   \item
	   University of Vienna, Department of Astrophysics, T\"urkenschanzstrasse 17, 1180 Wien, Austria
	   \item
	   Jodrell Bank Centre for Astrophysics, School of Physics \& Astronomy, University of Manchester, Manchester, UK
	   \item
	   Laboratory for Space Research, University of Hong Kong, Lung Fu Shan, Hong Kong
	   \item
	   Centre for mathematical Plasma-Astrophysics, KU Leuven, Celestijnenlaan 200B, 3001 Leuven, Belgium
\end{affiliations}
}


\begin{abstract}
In 1981, the idea of a superwind that ends the life of cool giant stars was proposed\cite{Renzini1981ASSL...88..431R}. Extreme OH/IR-stars develop superwinds with the highest mass-loss rates known so far, up to a few 10$^{-4}$\,\Msun/yr\cite{Heske1990A&A...239..173H,Delfosse1997A&A...320..249D,Chesneau2005A&A...435..563C, vanLoon2005A&A...438..273V,Justtanont2006A&A...450.1051J, Groenewegen2009A&A...506.1277G,Groenewegen2012A&A...543A..36G,Justtanont2013A&A...556A.101J,deVries2014A&A...561A..75D, Goldman2017MNRAS.465..403G,McDonald2018MNRAS.481.4984M}, hence epitomizing our understanding of the maximum mass-loss rate achieved during the Asymptotic Giant Branch (AGB) phase. A condundrum arises whereby the observationally determined duration of the superwind phase is too short for these stars to become white dwarfs \cite{Heske1990A&A...239..173H,Delfosse1997A&A...320..249D,Chesneau2005A&A...435..563C,Justtanont2006A&A...450.1051J,Groenewegen2012A&A...543A..36G,Justtanont2013A&A...556A.101J,deVries2014A&A...561A..75D}. Here, we report on the detection of spiral structures around two cornerstone extreme OH/IR-stars, OH\,26.5+0.6 and OH\,30.1$-$0.7, identifying them as wide binary systems. Hydrodynamical simulations show that the companion’s gravitational attraction creates an equatorial density enhancement mimicking a short extreme superwind phase, thereby solving the decades-old conundrum. This discovery restricts the maximum mass-loss rate of AGB stars around the single-scattering radiation-pressure limit of a few 10$^{-5}$\,\Msun/yr. This brings about crucial implications for nucleosynthetic yields, planet survival, and the wind-driving mechanism.
\end{abstract}

All stars with initial mass between 0.8-8 solar masses (\Msun) lose $\sim$40-80\% of their mass via a dust-driven wind during the Asymptotic Giant Branch (AGB) phase, their final destiny being white dwarfs. This wind is caused by radiation pressure on grains formed in the extended cool atmosphere due to pulsations\cite{Hofner2018A&ARv..26....1H}. For mass-loss rates above  $\sim$10$^{-7}$\,\Msun/yr, the associated timescale for stars to shed their mantle is much shorter than the nuclear burning timescale, such that mass loss determines the further evolution\cite{Hofner2018A&ARv..26....1H}. An empirical relation between mass-loss rate and luminosity is given by Reimers’ original formula\cite{Reimers1975MSRSL...8..369R} widely used in evolutionary calculations: \Mdot\,=\,$4\times10^{-13} \eta L/g R$, with \Mdot\ the mass-loss rate in units of solar masses per year (\Msun/yr), $\eta$ a unit-less parameter, and the stellar luminosity $L$, gravity $g$, and radius $R$ in solar units. For $\eta\sim1$, the AGB lifetime is of the order of one million years, and the maximum AGB mass-loss rate is only a few $10^{-6}$\,\Msun/yr\cite{Renzini1981ASSL...88..431R}. Hence, a Reimers-like wind cannot explain the distribution of white dwarf masses and characteristics of the planetary-nebulae population\cite{Renzini1981ASSL...88..431R}. Therefore, Renzini proposed in 1981 that a \textit{superwind} (see Supplementary Information for terminology) with mass-loss rate of at least a few $10^{-5}$\,\Msun/yr can develop at the high luminosity tip of the AGB phase leading to the ejection of the residual envelope on a timescale that is very short compared with the overall AGB lifetime. The debate over a sudden onset of the superwind phase or a gradual increase in mass-loss rate has not yet been settled, neither has the maximum mass-loss rate been established (see Supplementary Information).

Since 1981, many AGB stars were found to have mass-loss rates consistent with the suggested superwind values\cite{Knapp1985ApJ...292..640K,Bedijn1987A&A...186..136B,Wood1992ApJ...397..552W}. In this context, extreme OH/IR-stars are key objects. They represent a class of intermediate-mass (2--8\,\Msun) oxygen-rich stars at the tip of the AGB evolution with the \textit{highest} mass-loss rates discovered (between $10^{-5}$ and a few $10^{-4}$\,\Msun/yr) (see Supplementary Information). As such, these stars underpin a crucial part of our understanding of the maximum AGB mass-loss rate achieved, and hence of the process by which stars return mass to the interstellar medium. However, for more than a dozen extreme OH/IR-stars it has been shown that the superwind phase lasts only $\sim$200--1\,000 years and was preceded by a phase with mass-loss rate one or two orders of magnitude lower (around a few $10^{-6}$\,\Msun/yr)\cite{Heske1990A&A...239..173H,Delfosse1997A&A...320..249D,Chesneau2005A&A...435..563C,Justtanont2013A&A...556A.101J,deVries2014A&A...561A..75D}. Given an \textit{expected} superwind lifetime of some ten thousand years to allow the stars to reduce their mass below the Chandrasekhar limit, the probability of observing a dozen stars with almost the same superwind age (of $\sim$200--1\,000 years) is unrealistically small (see Methods). Statistical reasoning implies that the $\sim$10$^{-4}$\,\Msun/yr superwind duration for these extreme OH/IR-stars cannot be much longer than $\sim$1\,000 years (see Methods). This duration is, however, too short for these stars to become white dwarfs resulting in a conflict with initial-final mass relations deduced from young stellar clusters\cite{Catalan2008MNRAS.387.1693C}.

Two prototype stars of this class of extreme OH/IR-stars have been observed with ALMA. Observations of low-excitation rotational lines of $^{12}$CO were obtained for OH\,26.5+0.6 and OH\,30.1$-$0.7 (see Fig.~1, Fig.~2, see Methods, see Supplementary Information). The CO data show a centralized bright emission region with full width at half maximum (FWHM) of $\sim$0\farcs4--0\farcs6. This value is similar to the FWHM of the dusty envelope which was interpreted as a measure for the superwind envelope with duration of only $\sim$130--200 ($\sim$440--650) years for OH\,26.5+0.6 (OH\,30.1$-$0.7)\cite{Chesneau2005A&A...435..563C,Justtanont2013A&A...556A.101J,deVries2014A&A...561A..75D}. A distinct feature in the ALMA data of both targets is an incomplete ring-like pattern (see Fig.~1, Fig.~2). The position-velocity diagrams (see Fig.~\ref{Fig:PVs}) reveal that this incomplete ring-like pattern is characteristic for a \textit{shell-like spiral created by the orbital motion of the mass-losing AGB star around the center of mass of a binary system}\cite{Kim2012ApJ...759...59K,Homan2015A&A...579A.118H} (see Methods, see Supplementary Information). We deduce an orbital period of $\sim$430 ($\sim$590) years and mean binary separation of $\sim$24 ($\sim$38) stellar radii (\Rstar) for OH\,26.5+0.6 (OH\,30.1$-$0.7) (see Methods).

The detection of a binary-induced shell-like spiral in both OH/IR-stars puts the evolutionary and statistical problem of the short superwind phase in extreme OH/IR-stars in a new perspective. Hydrodynamical simulations for a binary system resembling OH\,26.5+0.6 (Fig.~15 in Mastrodemos et al.\cite{Mastrodemos1999ApJ...523..357M}) show a wind pattern similar to the ALMA data with the appearance of an incomplete ring-like pattern arising from a spiral shock with spiral-arm spacing (of $\sim$400\,\Rstar) comparable to the one derived for OH\,26.5+0.6 ($\sim$1\arcsec\ or $\sim$350\,\Rstar). Of prime importance for this study is the emergence of a compact high-density region within $\sim$150\,\Rstar\ from the primary star caused by the gravitational focusing by the binary companion (Fig.~15 in Mastrodemos et al.\cite{Mastrodemos1999ApJ...523..357M}). This extent of $\sim$150\,\Rstar\ corresponds to a radial crossing time of $\sim$100 years, much in resemblance with the duration of the high mass-loss (superwind) phase of OH\,26.5+0.6 ($\sim$130 years). This compact equatorial density enhancement (CEDE) has a density contrast of a factor $\sim$10 compared to the background intrinsic wind density\cite{Kim2012ApJ...759...59K} (see Methods). \textit{This CEDE renders superfluous the concept of a very short extreme superwind phase in OH\,26.5+0.6}, i.e.\ the compact higher density region in the vicinity of OH\,26.5+0.6 –-- and likewise in OH\,30.1$-$0.7 --– is caused by the direct gravitational interaction with a binary companion and does not signal an abrupt change in mass-loss rate. The ALMA position-velocity diagrams and asymmetries reported in other observational diagnostics allow us to deduce the orientation of the lower density orbital axis and higher density orbital plane (see Fig.~\ref{Fig:sketch}, see Methods).

The ALMA data of OH\,26.5+0.6 and OH\,30.1-0.7, and statistical reasoning imply a general argument that a large fraction, if not all, extreme OH/IR-stars reported to have a short $\sim$10$^{-4}$\,\Msun/yr superwind phase are part of a binary system. With an observed binary frequency for intermediate mass stars of at least 30--45\%, increasing to 60--100\% for extreme OH/IR stars with initial mass around 5--8\,\Msun\cite{Raghavan2010ApJS..190....1R,Duchene2013ARA&A..51..269D}, the chance of observing the imprints of binary interaction is high. As such, the intense superwind phase in extreme OH/IR-stars is neither short nor problematic. A CEDE resolves the statistically unlikely situation of detecting a dozen stars having the same short $\sim$10$^{-4}$\,\Msun/yr superwind age (of $\sim$200--1000 years) if the superwind lifetime would be much longer, as required from evolutionary considerations. Moreover, one cannot use the measured geometrical extent of the compact high-density region to estimate a ‘duration’ since the velocity vector field does not follow a radially-streaming pattern in the vicinity of the binary stars\cite{Mastrodemos1999ApJ...523..357M}. 

The idea of a binary-induced CEDE mimicking a short intense superwind phase also explains (i)~the continuous distribution of infrared colors of observed OH/IR samples, (ii)~the absence of evidence of episodic superwind bursts, (iii)~the bipolar or multipolar morphology frequently detected in (pre-)planetary nebulae (PNe), and (iv)~the detection of micron-sized grains in some (pre-)PNe (see Supplementary Information). An important consideration is that all extreme OH/IR-stars with suggested short intense superwind phase display crystalline silicate features. Since we attribute the density enhancement to binary interaction, and following argumentation on crystal formation\cite{Molster1999Natur.401..563M,Edgar2008ApJ...675L.101E}, we suggest processing of grains into crystalline lattice structures in the CEDE (see Methods). Conversely the presence of crystalline silicates in the envelope surrounding OH/IR-stars can be seen as an indicator for binary interaction creating an environment facilitating the crystallization of grains.

A crucial implication of this paradigm shift, linking the behavior of extreme OH/IR-stars to binary interaction, deals with stellar evolution, in particular with the \textit{maximum mass-loss rate} attained during the AGB phase. This rate rules the fate of the star and in a broader context dictates the chemical enrichment of the interstellar medium by evolved stars. Observational studies\cite{vanLoon1999A&A...351..559V} suggest that the maximum mass-loss rate during the superwind phase can exceed the single-scattering radiation pressure limit (\Mdot\,=\,$L/v c$, with $v$ the wind velocity, $c$ the speed of light, and typical values around a few $10^{-5}$\,\Msun/yr) by a factor of $\sim$10. Dynamical models put forward a more conservative factor around 1--2\cite{Bladh2015A&A...575A.105B,Schroder1999A&A...349..898S}. The realization that all reported high mass-loss rate values concern extreme OH/IR-stars (in binary systems) and are based on simplified 1D analyses of warm (crystalline) dust grains residing close to the star, hence in the CEDE, leads to the conclusion that the reported mass-loss rates have been overestimated with a factor of a few up to $\sim$100 (see Methods). Low-excitation CO lines are the prime accurate probes of the mass-loss rate yielding values around $\sim$(0.1--3)$\times10^{-5}$\,\Msun/yr (see Methods). \textit{This establishes a threshold for the maximum mass-loss rate around the single-scattering radiation pressure limit}. This conclusion has important consequences for stellar evolution models implementing mass-loss rate prescriptions (see Supplementary Information) and for our understanding of the wind-driven mechanism in AGB stars. Lowering the maximum mass-loss rate in stellar evolution models implies an increase in the AGB lifetime and impacts the nucleosynthetic yields returned to the interstellar medium both in terms of light and heavy (s-process) elements, due to the occurrence of extra thermal pulses and more third dredge-up episodes, with expected differences up to an order of magnitude\cite{Karakas2018MNRAS.477..421K}. A lower mass-loss rate over a longer time lapse also allows for a more progressive mass transfer and possibly a higher total AGB material captured by the companion in binary systems undergoing wind-mass transfer. This can explain why a fraction of carbon enhanced metal poor stars have accreted as much as 0.1--0.2 solar masses of material from their AGB companion\cite{Abate2016A&A...587A..50A}. A lower mass-loss rate also facilitates the survival of close-by gas planets by avoiding thermal evaporation\cite{Villaver2007ApJ...661.1192V}. State-of-the-art hydrodynamical models investigating the wind driving in oxygen-rich AGB stars based on stellar pulsations and radiation pressure on dust need substantial finetuning to reach mass-loss rates of $\sim$10$^{-5}$\,\Msun/yr\cite{Bladh2015A&A...575A.105B} and very speculative mechanisms have been invoked\cite{Szyszka2011MNRAS.416..715S} to explain mass-loss rates substantially beyond this limit. For a maximum mass-loss rate not exceeding the single-scattering radiation pressure limit, we infer that the dust-driven wind scenario can explain the mass loss in all AGB stars with pulsation periods above $\sim$300 days.




\begin{addendum}
 \item 
This paper uses the ALMA data ADS/JAO.ALMA2015.1.00054.S, 2016.1.00005.S, and 2016.2.00088.S. ALMA is a partnership of ESO (representing 
its member states), NSF (USA) and NINS (Japan), together with NRC 
(Canada) and NSC and ASIAA (Taiwan), in cooperation with the Republic of 
Chile. The Joint ALMA Observatory is operated by ESO, AUI/NRAO and NAOJ.
This paper makes use of the CASA data reduction package: \url{http://casa.nra.edu} -- Credit: International consortium of scientists based at the National
Radio Astronomical Observatory (NRAO), the European Southern Observatory (ESO), the National Astronomical Observatory of Japan (NAOJ), the CSIRO Australia Telescope National Facility (CSIRO/ATNF), and the Netherlands Institute for Radio Astronomy (ASTRON) under the guidance of NRAO.
L.D.,  T.D., W.H., and M.V.d.S.\ acknowledge support from the ERC consolidator grant 646758 AEROSOL, T.D.\ acknowledges support from the Fund of Scientific Research Flanders (FWO). D.A.G.H.\ acknowledges support provided by the Spanish Ministry of Economy and Competitiveness (MINECO) under grant AYA-2017-88254-P. We acknowledge the help of 
Carl Gottlieb (Harvard University, US) for his editorial advice on the manuscript.
\item[Author Contributions] L.D.\ identified the spiral structure in the ALMA data of OH\,26.5+0.6 and OH\,30.1$-$0.7, performed the full analysis, and lead the consortium, W.H., T.D., and A.d.K.\ contributed to the interpretation of the data, D.E., D.A.G.H.\ and S.M.\ proposed the ALMA observations (ALMA proposals 2015.1.00054.S, 2016.1.00005.S, and 2016.2.00088.S), S.M.\ reduced the ALMA data, D.E.\ did the sample analysis of the extreme OH/IR stars, G.M.\ give advice on the statistical matters, Y.E.M.\ ran the ballistic simulations, C.G.\ made Fig.~\ref{Fig:sketch}, all authors contributed to the discussion. 
\item[Author Information] Reprints and permissions information is available at www.nature.com/reprints.
 \item[Correspondence] Correspondence and requests for materials
should be addressed to Leen Decin~(email: Leen.Decin@kuleuven.be).
\item[Competing Interests] The authors declare that they have no competing financial interests.
 \item[Supplementary Information] is linked to the online version of the paper at www.nature.com/nature.
\end{addendum}

%
\begin{figure}[htp]
    \centering\includegraphics[width=.8\textwidth]{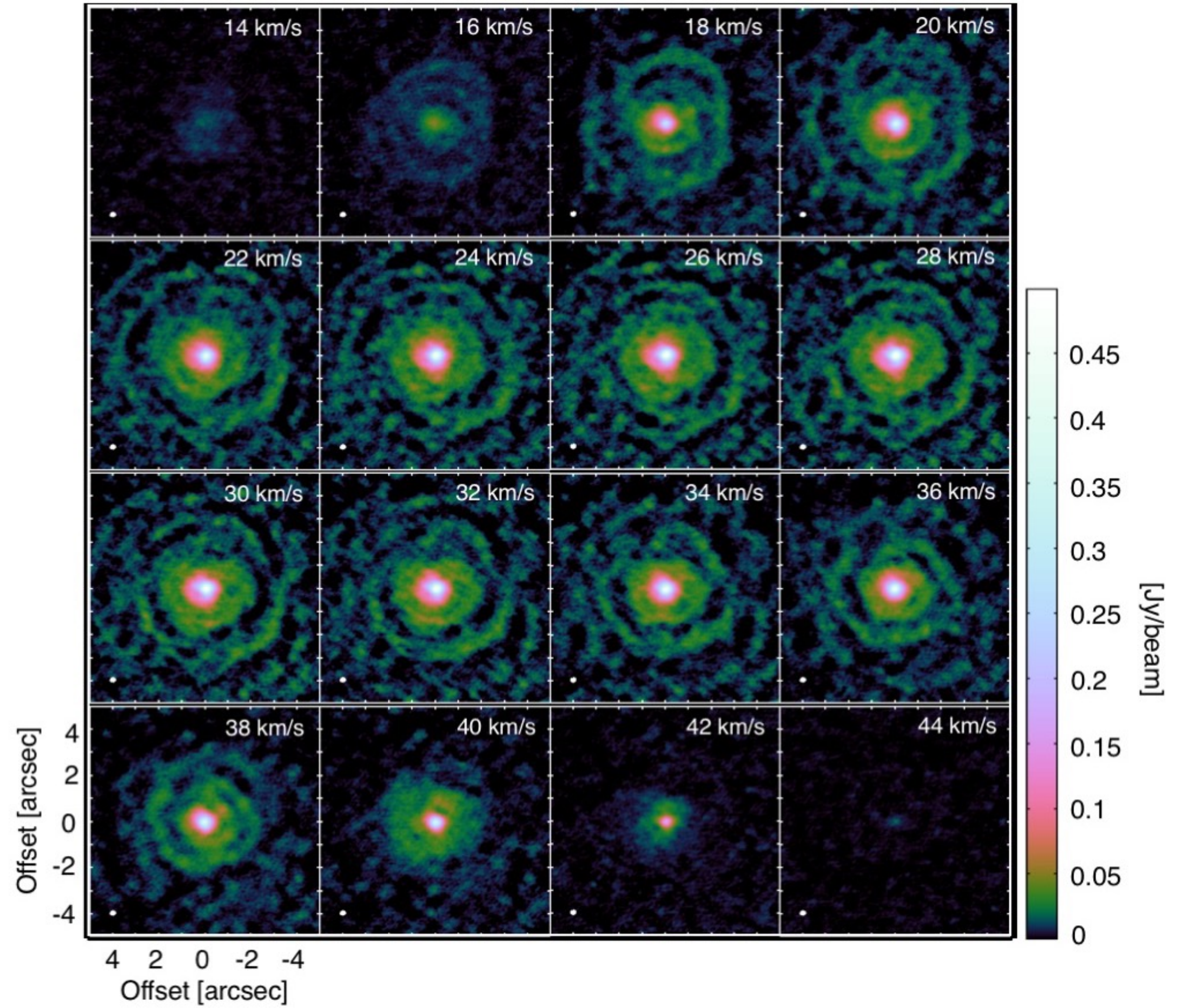}
\vspace*{-1ex}
    \caption{\textbf{ALMA $^{12}$CO J=3-2 channel map of OH\,26.5+0.6.} The map shows the incomplete ring-like pattern due to a binary-induced shell-like spiral structure. The ordinate and co-ordinate axis give the offset of the right ascension and declination, respectively. North is up, East is left. The ALMA beam size is shown as a white ellipse in the bottom left corner of each panel. The systemic velocity is at $\sim$28\,km/s. An animated video slicing through the velocity frames of the channel maps is provided in Supplementary Video~1. {\em The contrast is best visible on screen. }} 
   \label{Fig:channel_map_OH26}
  \end{figure}

%
\begin{figure}[htp]
    \centering\includegraphics[width=.8\textwidth]{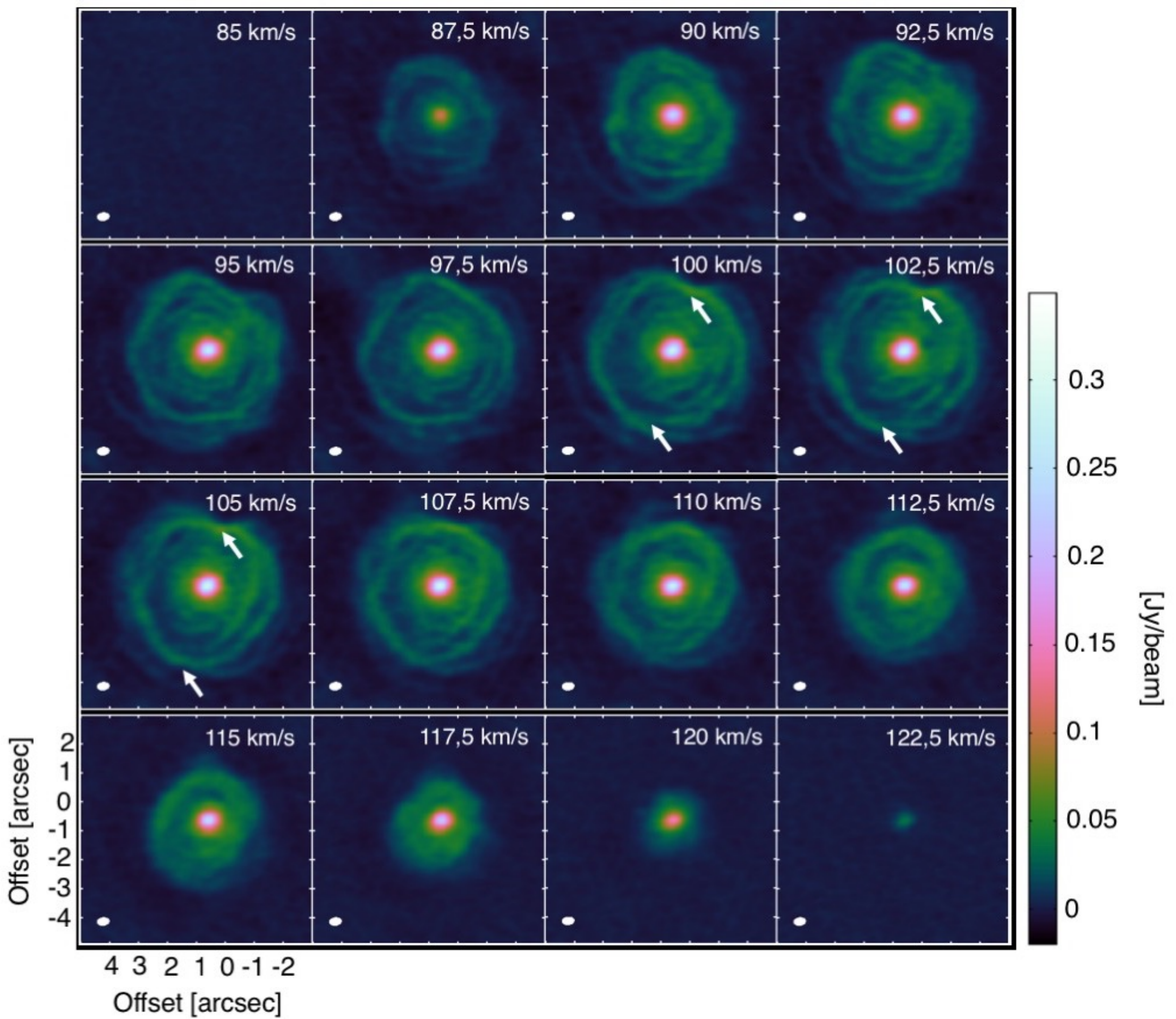}
\vspace*{-1ex}
    \caption{\textbf{ALMA $^{12}$CO J=3-2 channel map of OH\,30.1$-$0.7.} The map shows the incomplete ring-like pattern due to a binary-induced shell-like spiral structure. The white arrows indicate the knots caused by the overlap of gravitationally induced density wake by the companion with the spiral-shell pattern\cite{Kim2012ApJ...759...59K} (see Supplementary Information). The systemic velocity is at $\sim$99\,km/s. An animated video slicing through the velocity frames of the channel maps is provided in Supplementary Video~2. {\em The contrast is best visible on screen. }} 
    \label{Fig:channel_map_OH30}
 \end{figure}

\begin{figure}[htp]
\centering\includegraphics[width=10.5truecm]{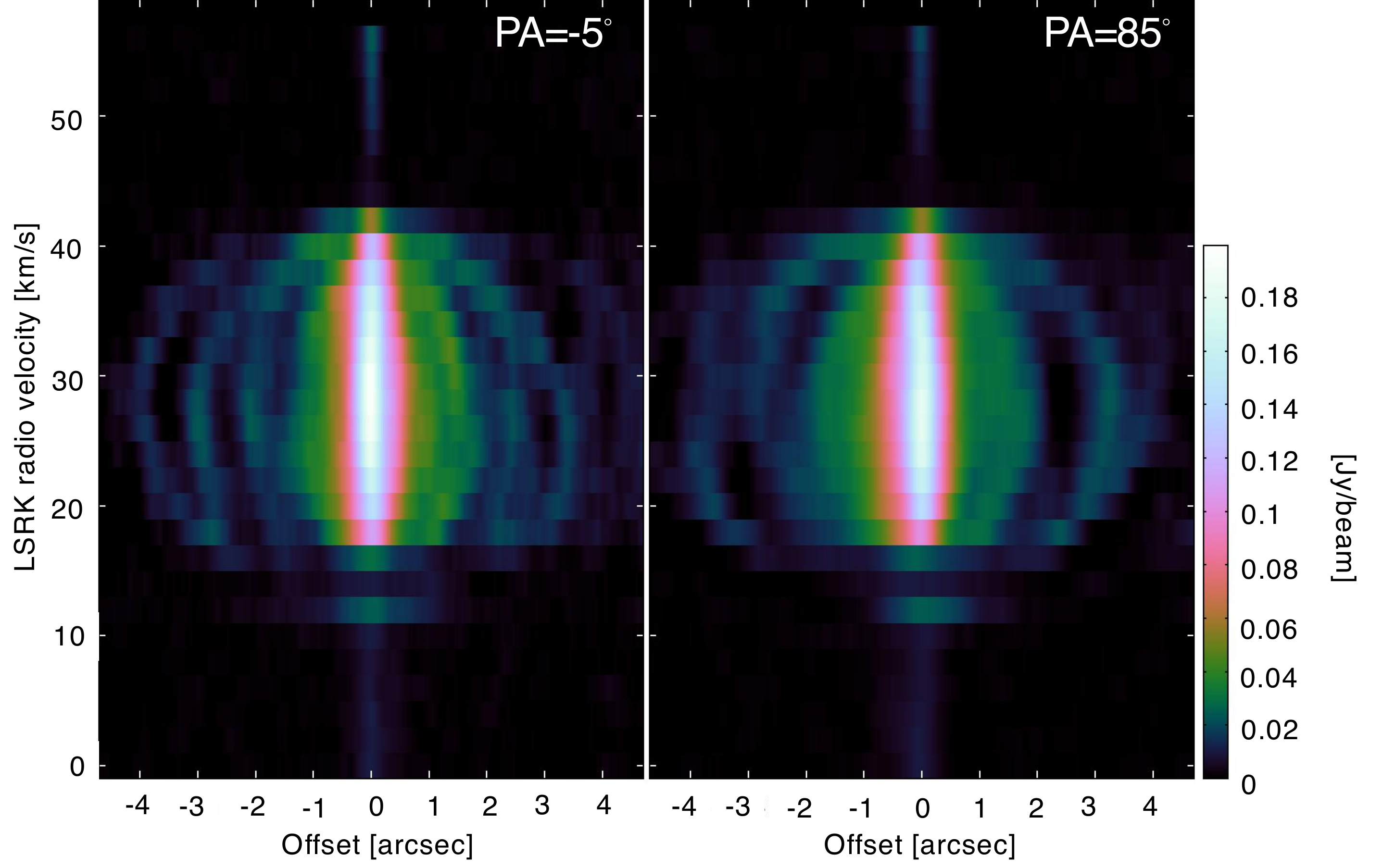}
\centering\includegraphics[width=11.5truecm]{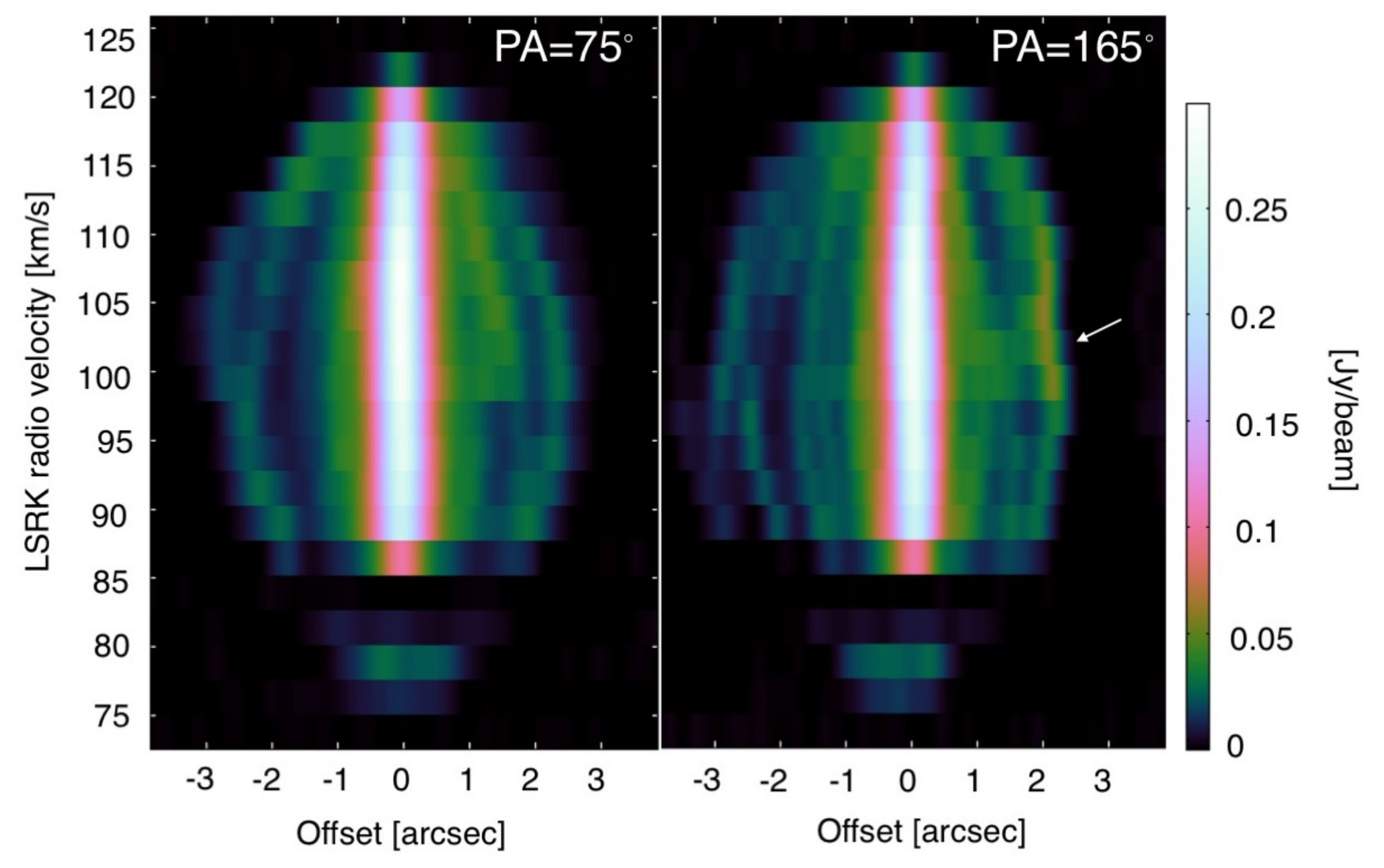}
\vspace*{-2.5ex}
\caption{\textbf{Position-velocity (PV) diagrams of the $^{12}$CO J\,=\,3--2 emission in OH\,26.5+0.6 (top) and of the the $^{12}$CO J\,=\,2--1 emission in OH\,30.1-0.7 (bottom).}
\textit{Top panel:} PV-diagram of the $^{12}$CO J\,=\,3--2 emission in OH\,26.5+0.6 for a position angle (PA) of $-$5$^\circ$ (left) and 85$^\circ$ (right).
The resulting PV-diagrams are characteristic for a {\em binary-induced shell-like spiral structure}\cite{Kim2012ApJ...759...59K,Homan2015A&A...579A.118H}. The binary system is viewed at an inclination angle, $i$, around 60$^\circ$--75$^\circ$ (see Fig.~11 and Fig.~13 in Homan et al.\cite{Homan2015A&A...579A.118H}). The systemic velocity is at $\sim$28\,km/s. The lack of emission at velocities between 12\,--\,16\,km/s is caused by the blue-wing absorption\cite{Decin2018arXiv180109291D}.
\textit{Bottom panel:} PV-diagram of the $^{12}$CO J\,=\,2--1 emission in OH\,30.1-0.7 for a position angle (PA) of 75$^\circ$ (left) and 165$^\circ$ (right).
As in the case of OH\,26.5+0.6, the resulting PV-diagrams are characteristic for a {\em binary-induced shell-like spiral structure}\cite{Kim2012ApJ...759...59K,Homan2015A&A...579A.118H}. In the PV-diagram along the line of nodes of the knots (see Fig.~\ref{Fig:channel_map_OH30}), hence at a PA of 165$^\circ$, the knotty structure is present at the systemic velocity (of $\sim$99\,km/s, see white arrow), confirming the existence of the companion’s wake in the orbital plane\cite{Kim2013ApJ...776...86K}.
The binary system is viewed at an inclination angle, $i$, around 50$^\circ$--70$^\circ$.  The lack of emission at velocities between 80\,--\,85\,km/s is caused by the blue-wing absorption\cite{Decin2018arXiv180109291D}.
 {\em The contrast is best visible on screen.}}
 \label{Fig:PVs}
\end{figure}

\begin{figure}[htp]
\centering\includegraphics[width=12truecm]{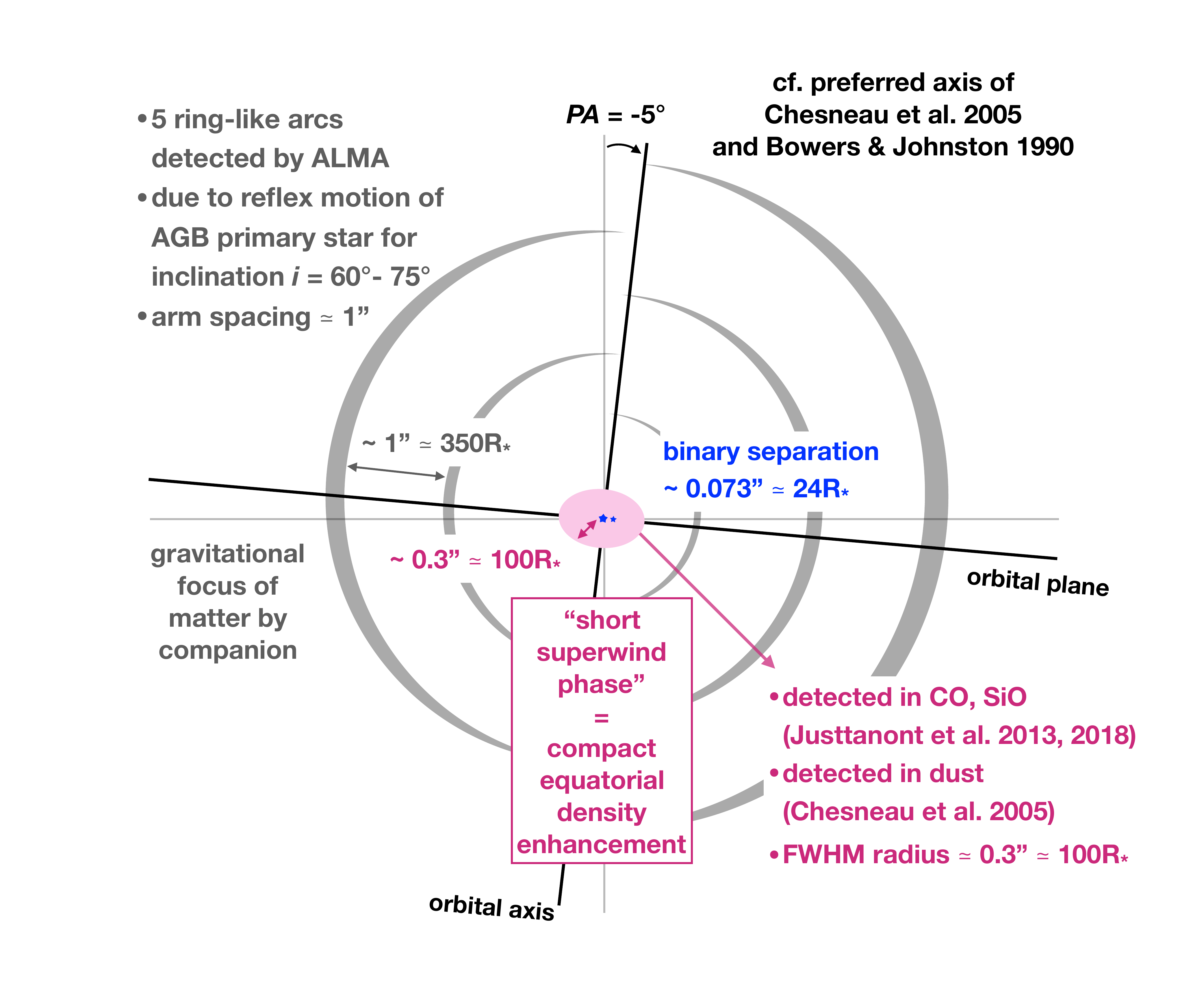}
\caption{\textbf{Sketch of the \oh\ binary system.} Sketch of
the inner 10\arcsec\ wind region of \oh\ indicating the different observations that signal a binary companion. The reflect motion of the primary OH/IR-star causes a one-arm spiral structure with an extent almost reaching the orbital axis. This spiral structure is seen at a high inclination angle ($i$\,=\,60$^\circ$\,--\,75$^\circ$) and as such reveals itself in an incomplete ring-like pattern (see grey areas in the sketch). From the ALMA data, we infer an orbital axis with a position angle (PA) of $-5^\circ\pm10^\circ$.
A higher density region with FWHM of $\sim0\farcs6$ (in pink) is detected in molecular (CO, SiO) and dust emission. This region was hitherto misinterpreted as a sign for a very short superwind phase. We show that this region is an equatorial density enhancement situated in the orbital plane  caused by the direct gravitational focussing of the binary companion.}
\label{Fig:sketch}
\end{figure}



\afterpage{\clearpage}
\newpage

\begin{methods}


\section*{The statistical probability of observing stars having a $\sim$10$^{-4}$\,\Msun/yr superwind with age of $\sim$1\,000 years} \label{Methods:statistics}

For extreme OH/IR-stars (see Supplementary Information for the definition of this class of objects) like \oh\ and \OH30\ with an initial mass between 4--8 solar masses, the estimated Asymptotic Giant Branch (AGB) lifetime during the thermally pulsating phase is between 0.01--0.35 million years\cite{Karakas2014MNRAS.445..347K}. More than a dozen extreme OH/IR stars within a distance of $\sim$4\,0000 parsec have been shown to have a current $\sim$10$^{-4}$\,\Msun/yr superwind that was preceded by a phase with mass-loss rate one to two orders of magnitude lower (mean mass-loss rate $\sim$7.4$\times$10$^{-6}$\,\Msun/yr)\cite{Heske1990A&A...239..173H,Delfosse1997A&A...320..249D,Chesneau2005A&A...435..563C,Justtanont2006A&A...450.1051J,Groenewegen2012A&A...543A..36G,Justtanont2013A&A...556A.101J,deVries2014A&A...561A..75D} (see Supplementary Information). Modelling suggests that the onset of the present-day high mass-loss rate phase was only $\sim$200--1\,000 years ago\cite{Heske1990A&A...239..173H,Delfosse1997A&A...320..249D,Chesneau2005A&A...435..563C,Justtanont2006A&A...450.1051J,Groenewegen2012A&A...543A..36G,Justtanont2013A&A...556A.101J,deVries2014A&A...561A..75D}. For a superwind with mass-loss rate of $\sim$10$^{-4}$\,\Msun/yr a superwind lifetime of some ten thousand years is required so that these stars lose enough mass to reach white dwarf masses. 
The probability of observing a dozen stars with almost the {\em same superwind age} (of 200\,--\,1\,000 years) is extremely small if the lifetime of the superwind is of the order of ten thousand years. This can easily be derived by defining
\begin{itemize}
\setlength{\itemsep}{-1mm}
\item $N$: the number of stars in the population,
\item $n$: the number of stars observed; obviously $n \le N$,
\item $p$: the number of `successful' stars, i.e.\ stars within the window specified, in the case having a $\sim$10$^{-4}$\,\Msun/yr superwind with age of $\sim$1\,000 years; obviously $0\le p\le n$,
\item $\pi$: probability that a star falls within the specified window; for this problem
$\pi\sim 1\,000/10\,000=0.1.$
\end{itemize}
Assuming observations are independent of each other, the probability of observing $p$ `successes' out of $n$ is binomial:
$$P(p|n,\pi)=\left(\begin{array}{c}n\\p\end{array}\right)\pi^p(1-\pi)^{n-p}.$$
One can roughly estimate the number of AGB stars which are type~II OH/IR-stars (see Supplementary Information for terminology) and with infrared colours similar to \oh\ and \OH30, using the space densities derived by Ortiz et al.\cite{Ortiz1996A&A...313..180O}. This yields $\sim$400 stars within a sphere of radius $\sim$4\,000 parsec around the Sun, hence $N\sim400$. 
For a dozen of extreme OH/IR-stars, all necessary (spectroscopic) data of the infrared dust and CO emission are available to estimate the age of the superwind. For {\em all} these galactic extreme OH/IR-stars, a short superwind age was retrieved\cite{Heske1990A&A...239..173H,Delfosse1997A&A...320..249D,Chesneau2005A&A...435..563C,Justtanont2006A&A...450.1051J,Groenewegen2012A&A...543A..36G,Justtanont2013A&A...556A.101J,deVries2014A&A...561A..75D}. 
Hence, $p=n\approx12$ or\\
$$P(p=12|n=12,\pi)=\left(\begin{array}{c}12\\12\end{array}\right)\pi^{12}(1-\pi)^{0}=\pi^{12}=10^{-12}.$$
This low probability implies that the superwind lifetime in intermediate-mass stars cannot be of the order of ten thousand years. This is confirmed by the non-detection of extreme OH/IR-stars with a $\sim$10$^{-4}$\,\Msun/yr superwind with age between $\sim$2\,000 and 10\,000 years. Under the assumption of single-star evolution (but see main text), the detection of {\em all observed} extreme OH/IR-stars in that extremely short evolutionary phase can statistically only be understood if the duration of the $\sim$10$^{-4}$\,\Msun/yr superwind phase is not much longer than $\sim$1\,000 years (hence $\pi$ approaches 1). However, this duration is too short for these stars to become white dwarfs resulting in a conflict with initial-final mass relations deduced from young stellar clusters\cite{Catalan2008MNRAS.387.1693C}.

\section*{ALMA observations of \oh\ and \OH30} \label{methods:ALMA_data}

Aiming to better quantify the short superwind phase of \oh\ and \OH30, low-excitation $^{12}$CO lines of both targets have been observed with ALMA (see Supplementary Information). The data were reduced using the CASA software package\cite{CASA2007ASPC..376..127M}. The full potential of the ALMA $^{12}$CO data is laid out in the channel maps (see Fig.~\ref{Fig:channel_map_OH26}, Fig.~\ref{Fig:channel_map_OH30}).  
 Both the channel maps and zero-moment maps (see Supplementary Fig.~1, Supplementary Fig.~2 show bright emission in the central region with full width half maximum (FWHM) of $\sim$0\farcs4--0\farcs6. For a spherically symmetric wind, this corresponds to a duration of $\sim$130--200 ($\sim$440--650) years for \oh\ (\OH30) being similar to the superwind age derived for both targets (see Supplementary Information). 
 
Of particular interest for this study is the emergence of an {\em incomplete ring-like pattern} in the $^{12}$CO channel map of both targets (see Fig.~\ref{Fig:channel_map_OH26}, Fig.~\ref{Fig:channel_map_OH30}, see Supplementary Information). This ring-like pattern signals the presence of a binary companion\cite{Kim2012ApJ...759...59K,Maercker2012Natur.490..232M,Kim2013ApJ...776...86K,Decin2015A&A...574A...5D,Homan2015A&A...579A.118H}. Position-velocity (PV) diagrams allow one to gauge if and how density structures are correlated.  The PV diagrams prove that the circumstellar envelope of both targets is shaped by a \textit{binary-induced shell-like spiral}\cite{Kim2012ApJ...759...59K,Homan2015A&A...579A.118H} created by the mass-losing AGB star wobbling around the center-of-mass of a binary system (see Fig.~\ref{Fig:PVs}, see Supplementary Information).   In addition, the high signal-to-noise data of \OH30\ allow us to detects knots (see white arrows in Fig.~\ref{Fig:channel_map_OH30}) which indicate the region where the shell-like spiral pattern overlaps with the density wake caused by Bondi-Hoyle accretion of the binary companion (see Supplementary Information).

\section*{Binary properties of \oh\ and \OH30}


Hydrodynamical simulations show that for a mass-losing star in a binary system wobbling around the center-of-mass in a circular orbit an Archimedean spiral is formed in the circumstellar envelope\cite{Kim2012ApJ...759...59K}. If the orbital plane is located exactly on the plane of the sky ($i=0^\circ$), this Archimedean spiral can easily be recognised in an angle-radius plot as straight lines (see upper panel in Supplementary Fig.~3). Including the effect of inclination, a spiral-shell structure exhibits undulations with respect to these straight lines\cite{Kim2013ApJ...776...86K}. This is illustrated for the \oh\ binary system in the bottom panel of Supplementary Fig.~3.
As for CIT~6\cite{Kim2013ApJ...776...86K}, the inclination effect of the \oh\ binary system can easily be discerned in Supplementary Fig.~3. 
Exploiting the asymmetry in the PV-diagrams and the presence of knots in the \OH30\ data, one can use the outcome of radiative transfer calculations\cite{Kim2013ApJ...776...86K,Homan2015A&A...579A.118H} to estimate the inclination angle, $i$, of the binary system which is between $\sim$60$^\circ$\,--\,75$^\circ$ for \oh, and between $\sim$50$^\circ$\,--\,70$^\circ$ for \OH30.

Assuming a circular orbit, the orbital period can be estimated from the slope of the observed emission in the angle-radius plot\cite{Kim2017NatAs...1E..60K}, which for \oh\ yields $\sim$430 years. Another method exploits the spiral-arm spacing\cite{Kim2012ApJ...759L..22K,Decin2015A&A...574A...5D}. The spiral-arm spacing is determined by the product of the binary orbital period $T_p$ and the pattern propagation speed in the orbital plane, which is given by\cite{Kim2012ApJ...759L..22K}
\begin{equation}
 \Delta r_{\rm{arm}} = \Big(\langle V_w \rangle + \frac{2}{3} V_p \Big) \times \frac{2 \pi r_p}{V_p}\,,
 \label{Eq:Tp}  
\end{equation}
with $\langle V_w \rangle$ the wind velocity, $V_p$ the orbital velocity, $r_p$ the orbital radius of the primary, and the orbital period $T_p$ given by the last term ($ 2 \pi r_p/V_p$). The first term in the right-hand part of Eq.~\ref{Eq:Tp}, $\langle V_w \rangle + \frac{2}{3} V_p$, is the pattern propagation speed throughout the orbital plane. In AGB binary simulations, this pattern speed is close to the wind speed\cite{Kim2012ApJ...759...59K}. At the distance where the spiral arcs are detected (beyond $1\farcs5$ for \oh\ and $1\farcs0$ for \OH30, corresponding to $\sim$500 and 1\,000 stellar radii, respectively), the wind has already reached its terminal velocity. 
The derived arm-spacing of $\sim$1\arcsec\ for \oh\ and of $\sim0\farcs58$ for \OH30\ yields an orbital period of $\sim$350--430 years for \oh\ and of $\sim$460--590 years for \OH30,
assuming orbital speeds ranging up to 5\,km/s. Orbital periods for other (carbon-rich) AGB binary stars displaying spiral structures range from 55\,--\,830 years\cite{Mauron2006A&A...452..257M,Dinh2009ApJ...701..292D,Maercker2012Natur.490..232M,Kim2013ApJ...776...86K,Decin2015A&A...574A...5D}. 

Isotopic ratios of carbon and oxygen indicate a progenitor mass for both targets between 5 to 8\,\Msun\cite{Justtanont2015A&A...578A.115J} (see Supplementary Information). The current ALMA data do not allow an accurate estimate of the companion's mass. Assuming a total mass for the binary system between 5 to 9\,\Msun\ and using Kepler's third law, the mean binary separation is 100$\pm$20 astronomical units for \oh\ and 123$\pm$22 astronomical units for \OH30. For a stellar radius of $\sim$900 solar radii (\oh) and $\sim$700 solar radii (\OH30)\cite{DeBeck2010A&A...523A..18D}, this corresponds to $\sim$24 stellar radii and $\sim$38 stellar radii, respectively.

Currently, we know of some dozen extreme OH/IR stars within $\sim$4\,000 parsec having similar mass-loss rate characteristics as \oh\ and \OH30. Statistical reasoning implies a general argument that a large fraction of them is in a binary system. Using the space densities of Ortiz et al.\cite{Ortiz1996A&A...313..180O}, we expect some 400 stars within a $\sim$4\,000 parsec sphere having infrared colours similar to \oh\ and \OH30. The frequency for binary companions with a mass ratio of 0.1 to 1.0 and an orbital period ranging from $\sim$100 -– 1000 years is $\sim$13\% for stars with mass between 2 and 5\,\Msun\ and is $\sim$17\% for stars between 5 and 9\,\Msun\cite{Moe2017ApJS..230...15M}. Only considering mass ratios between 0.1 and 0.3, the binary frequency reduces to $\sim$6\% and $\sim$9\%, respectively. As such, we expect more extreme OH/IR stars to be detected being part of a binary system.

\section*{Properties of the compact equatorial density enhancement (CEDE)}

The seminal work of Mastrodemos \& Morris\cite{Mastrodemos1999ApJ...523..357M} is one of the few studies exploring the effects of the hydrodynamical interaction in AGB binaries for a large parameter space of wind velocities, circular orbit separations, and companion masses. Although their simulations were not fine-tuned toward the specific situation of \oh\ or \OH30\ (i.e.\ the estimated mass of the OH/IR-stars is larger than the adopted mass of the primary of 1.5 solar masses), their results can be used to gauge the family of wind patterns that can arise. Of particular interest for this study is their model M7 which has a large binary period of 227 years, a binary separation of 24 stellar radii, and a wind velocity of 15\,km/s, very similar to the deduced binary parameters of \oh. The modelled wind pattern perpendicular to the orbital plane clearly shows the appearance of an incomplete ring pattern arising from a spiral shock (see Fig.~15 in Mastrodemos \& Morris\cite{Mastrodemos1999ApJ...523..357M}). The arm-spacing is around 400 stellar radii, which in the case of \oh\ would translate to $\sim1\farcs2$, comparable to the one derived from the ALMA data. Of prime importance for this study is the emergence of a compact high-density region within $\sim$150 stellar radii (hence $\sim$6 times the binary separation) from the primary star caused by the gravitational focussing of the binary companion\cite{Mastrodemos1999ApJ...523..357M}.

A lower limit for the density enhancement of the compact equatorial density enhancement (CEDE) can be obtained using ballistic trajectory calculations\cite{ElMellah2017MNRAS.467.2585E}. Doing so, we have calculated the density amplification factor defined as the invert of the fraction of solid angle covered by the wind:
\begin{equation}
1/\sin\left(\theta_c\right)=1/\sin\left[\arctan\left(H/r\right)\right]
\end{equation}
where $H$ is half of the extent of the wind in the direction normal to the orbital plane, $r$ is the distance at which $H$ is measured and $\theta_c$ is the angle subtended (see top panel in Supplementary Fig.~4). A small grid of simulations with binary parameters similar to the ones of OH\,26.5$+$0.6 and OH\,30.1$-$0.7 led to the estimates in the bottom panel of Supplementary Fig.~4, where the binary mass ratio has been set to 5 and the filling factor of the donor star to 10\%. The variable $\beta$ represents how fast the wind reaches its terminal speed using a $\beta$-type velocity law\cite{Lamers1999isw..book.....L}: for $\beta=2$ (resp.\ $\beta=3$), the wind reaches 90\% of its terminal wind speed after 20 stellar radii (resp.\ 30 stellar radii). The beaming toward the equator is strongly dependent on $\eta$, being the ratio of the terminal wind speed to the orbital speed, and on $\beta$. These simulations show a density amplification of a factor of a few, which is a lower limit since cooling, which is not accounted for, will compress the wind even more\cite{ElMellah2018arXiv181012933E}.

The hydrodynamical simulations of Kim \& Taam\cite{Kim2012ApJ...759...59K} include all physics to better quantify the density enhancement factor. Using their simulation results, one can determine that the CEDE has a density contrast of a factor $\sim$10 with respect to the background wind density. This can be deduced by comparing Fig.~4 and Fig.~10 of Kim \& Taam\cite{Kim2012ApJ...759...59K}. Fig.~4 only includes the effect of the wobbling of the mass-loss AGB star around the center-of-mass (model M6), while in Fig.~10 both the wobbling of the mass-loss AGB star and the gravitational focusing by the secondary are accounted for (model M7). The density of model M7 is a linear superposition of the amounts of density enhancement due to the two mechanisms. The difference in density in the inner regions between Fig.~4 and Fig.~10 is roughly a factor of 10. The density enhancement increases for shorter radial distances and is strongly latitude-dependent. We note, however,  that values between a factor of a few up to $\sim$100 are not unusual depending on the specific parameters of the binary system\cite{Kim2012ApJ...759...59K,Mastrodemos1999ApJ...523..357M,Edgar2008ApJ...675L.101E,Liu2017ApJ...846..117L,Chen2017MNRAS.468.4465C}.


The bright centralized emission seen in the $^{12}$CO ALMA data of \oh\ and \OH30\ signals the CEDE (Fig.~\ref{Fig:channel_map_OH26}, Fig.~\ref{Fig:channel_map_OH30}, Supplementary Video~1, Supplementary Video~2). 
The creation of a CEDE in \oh\ is also supported by the infrared image of the dusty envelope of \oh\ that displays a slightly oblate shape with a PA for the major axis being at 95$^\circ \pm$6$^\circ$\cite{Chesneau2005A&A...435..563C} . The infrared asymmetry is also correlated with the asymmetry reported for the OH 1612\,MHz maser for which a projected axis of rotation oriented in the north-south direction was deduced\cite{Bowers1990ApJ...354..676B,Etoka2010MNRAS.406.2218E}. As such, both the ALMA PV diagrams of \oh\ and the oblateness of the OH and infrared dust images show morphological evidence for the north-south direction corresponding to the lower density orbital axis and the east-west direction to the higher density orbital plane (see Fig.~\ref{Fig:sketch}).
The direction of the orbital plane of \OH30\ is more difficult to constrain due to a lack of complementary observations. At the other hand, the presence of knots in the channel map (Fig.~\ref{Fig:channel_map_OH30}) is a strong diagnostic. In the PV-diagram at a PA of 165$^\circ$, being along the line of nodes of the knots,  the knotty structure is present at the systemic velocity (see Fig.~\ref{Fig:PVs}), confirming the existence of the companion’s wake in the orbital plane\cite{Kim2013ApJ...776...86K}.

This CEDE also offers the perfect conditions for amorphous grains to be annealed into crystalline minerals. As shown in the simulations of Edgar et al.\cite{Edgar2008ApJ...675L.101E}, the local deflection of the wind flow by the secondary drives a shock formation lying in the orbital plane  with an associated increase in temperature. A first estimate of the post-shock temperature can be calculated using Eq.~2 of Edgar et al.\cite{Edgar2008ApJ...675L.101E}. For the \oh\ and \OH30\ binary system parameters derived above, it is straightforward to calculate that post-shock temperatures between 1067\,K (the annealing temperature of silicates\cite{Hallenbeck2000ApJ...535..247H}) and 2000\,K (when grains are vaporized) are reached for the primary mass being between 6.3\,\Msun\ and 8.8\,\Msun, in very good agreement with the mass estimates from the isotope analysis\cite{Justtanont2015A&A...578A.115J}. The associated timescale for annealing is so short compared to other timescales\cite{Edgar2008ApJ...675L.101E} that the annealing is almost instantaneous. Hence, the shaping of the AGB wind by a binary companion provides a mechanism for forming crystalline dust in the orbital plane.

\section*{Derived mass-loss rates for extreme OH/IR-stars}

While mass-loss rates for OH/IR-stars are sometimes derived from the OH maser intensity\cite{Baud1983A&A...127...73B,vanderVeen1989A&A...226..183V}, the uncertainties on the derived values is large due to the inherent non-linear behaviour of masers\cite{Marschall2004MNRAS.355.1348M}. Therefore, the methods most frequently used for mass-loss rate determinations are the analyses of (i) infrared colours and/or dust spectral features or (ii) rotational transitions of CO gas. While both methods often suffer from the same simplification of a 1D (spherically symmetric) geometry, the first method involves the unknown dust-to-gas ratio and the unknown dust outflow velocities. The latter are often assumed to equal the gas outflow velocities, which is a valid assumption in the case of gas and dust being well coupled.
For the context of this paper, one needs to assess how the binary-induced morphology of the circumstellar wind, in particular the spiral structure and the compact equatorial enhancement (CEDE), complicates the derivation of the mass-loss rate.

\subsection{Tracing the mass-loss rate: warm dust versus cool gas:}

The mass-loss rates derived from the analyses of dust spectral features (or infrared colours) and from low-excitation CO lines yield significantly different results for extreme OH/IR-stars  with the former ones  being one or two orders of magnitude larger than the values derived from CO gas (see Supplementary Information). For \oh\ and \OH30, the values are  2--3$\times$10$^{-4}$\,\Msun/yr from dust versus $\sim$1\,--\,9.7$\times$10$^{-6}$\,\Msun/yr from CO \cite{Heske1990A&A...239..173H,DeBeck2010A&A...523A..18D,Justtanont1996ApJ...456..337J,Justtanont2013A&A...556A.101J,deVries2014A&A...561A..75D}, with the single-scattering radiation pressure limit being around 5$\times$10$^{-5}$\,\Msun/yr. This discrepancy was historically interpreted as a sign of a short intense superwind phase in which the (warm) dust traces the recent high density material close to the star, while the (cool) gas signals a lower mass-loss rate in the past (see Supplementary Information). However, realising that \oh\ and \OH30\ are part of a binary system in which both a compact equatorial density enhancement (CEDE) and spiral structure are formed puts this discrepancy in another context. I.e., the (near-)infrared spectral energy distribution (SED) mainly traces the {\em warm (crystalline) dust} residing close to the star, hence in the CEDE. The density in the CEDE is roughly a factor 10 higher than the background wind density. Therefore, the reported high mass-loss rate superwind values are mainly a reflection of the higher density in the CEDE created by binary interaction, but not of the actual mass-loss rate which will be lower. An additional complication is that SED fits for extreme OH/IR-stars have always been based on a simplified 1D geometry. This assumption might lead to an erroneous derivation of the density structure (see Supplementary Information). Moreover, with the velocity vector field being highly complex in the inner wind region due to the binary interaction \cite{Mastrodemos1999ApJ...523..357M}, a simple relation between density structure and mass-loss rate is ruled out.

The real mass-loss rate is better estimated from radiative transfer modelling of various low-excitation CO rotational lines observed with telescopes having a large beam size (hence incorporating the total wind extent). The advantage of low-excitation CO lines is that they probe the bulk of the material in the cooler less dense region of a stellar wind. As such, the mass in the CEDE is negligible compared to the mass probed. 
One might wonder if \textit{(i.)}~the occurrence of a large spiral structure, \textit{(ii.)} the change in mass-loss rate during the AGB evolution, or \textit{(iii.}) CO line saturation effects
do not complicate the mass-loss rate retrieval from low-excitation CO gas. Concerning the first aspect, we can rely on the studies performed by Homan et al.\cite{Homan2015A&A...579A.118H}. They have shown (see their Fig.\ 16, upper left panel) that one can safely analyse low-excitation CO emission lines using a simplified 1D approach in the case of a high mass-loss rate wind in which a shell-like spiral is embedded. To constrain the second aspect, we turn our attention again to \oh\ and \OH30, since both targets have been studied with various instruments. 
For \oh, the CO J=1--0 line was imaged with BIMA and a CO envelope diameter of 8.5\arcsec$\times$5.5\arcsec\ (FWHM) was derived\cite{Fong2002A&A...396..581F}. For a distance of 1370 parsec and a wind velocity of 15\,km/s (see Supplementary Information), this translates to $\sim$2600 years, which is $\sim$25\% of the interpulse period for a star with initial mass of 5\,\Msun\cite{Karakas2014MNRAS.445..347K}. Analogously, the $^{12}$CO J=1--0 data of \OH30\  yield an extent of only $\sim$3780 years (see Supplementary Information).
For stars with initial mass above $\sim$1.5\,\Msun, the mass-loss rate at the high-luminosity tip of the AGB will only marginally change during that time interval\cite{Schroder1999A&A...349..898S}, and hence the assumption of a constant mass-loss rate when analysing a set of low-excitation CO lines is valid. 
The third aspect occurs for stars experiencing a high mass-loss rate, since the saturation of the optically thick CO lines weakens the dependence of the line intensity on the amount of CO in the envelope\cite{Ramstedt2008A&A...487..645R}. This effect starts to play a role for mass-loss rates above $\sim$3$\times$10$^{-5}$\,\Msun/yr (i.e., similar to the single-scattering pressure limit for extreme OH/IR-stars), and the application of analytical formulae relating the CO line intensity with mass-loss rate should be done with care\cite{DeBeck2010A&A...523A..18D}. For mass-loss rates above that limit, dedicated radiative transfer modelling can cope with that effect\cite{Decin2006A&A...456..549D}. Hence, none of the three aforementioned aspects impedes the use of low-excitation CO lines to derive the mass-loss rate of (extreme) OH/IR-stars.
Additional advantages of the use of low-excitation CO lines are that the dust-to-gas conversion ratio is not needed and that the width of the CO lines yields the overall wind expansion velocity since the spiral-pattern propagation speed (Eq.~\ref{Eq:Tp}) is close to the wind speed\cite{Kim2012ApJ...759...59K}. Therefore, the equation of mass-conservation, \Mdot = $4 \pi r^2 \rho(r) v(r)$, can be used to convert the derived gas density structure, $\rho(r)$, into a value for the mass-loss rate. 
Hence, we conclude that various low-excitation CO lines should be used to determine the mass-loss rate in (extreme) OH/IR-stars yielding typical values around (0.1--3)$\times$10$^{-5}$\,\Msun/yr\cite{Heske1990A&A...239..173H,Justtanont1996ApJ...456..337J,Justtanont2013A&A...556A.101J}, i.e.\ \textit{not exceeding the single-scattering radiation pressure limit, and potentially even remaining lower by a factor of a few}. 

The uncertainty on mass-loss rate values as derived from low-excitation CO lines is typically a factor of $\sim$3\cite{Ramstedt2008A&A...487..645R}, which is partly caused by the impossibility to have direct knowledge on the [CO/H$_2$]-ratio. The latter value is taken from AGB stellar nucleosynthesis models which predict a value around $3\times10^{-4}$ with an uncertainty of a factor of 2\cite{Knapp1985ApJ...292..640K,Karakas2010MNRAS.403.1413K,Cristallo2015ApJS..219...40C}. It might be expected that the mass-loss rate varies during thermal pulse cycles. Following the argumentation given in the Supplementary Information based on dynamical atmosphere models\cite{Schroder1999A&A...349..898S}, we expect that as the star evolves and the mass-loss rate reaches the single-scattering radiation pressure limit, the impact of thermal pulses or pulsations on variations in the density structure is modest\cite{Vassiliadis1993ApJ...413..641V}, being a factor $\sim$2. Taking this into account and realizing that the single-scattering limit for extreme OH/IR stars is $\sim$5$\times$10$^{-5}$\,\Msun/yr, this study puts forward a maximum mas-loss rate during the AGB-phase around the single-scattering radiation pressure limit.

\begin{addendum}
\item[Data availability] 
The ALMA data from proposals 2015.1.00054.S, 2016.1.00005.S, and 2016.2.00088.S can be retrieved from the ALMA data archive at \url{http://almascience.eso.org/aq/}. The data that support the plots within this paper and other findings of this study are available from the corresponding author upon reasonable request.
\item[Code availability] Figs.~1--3, and Supplementary Figs.~1--2 are made using CASA\cite{CASA2007ASPC..376..127M}. Supplementary Fig.~3 is made using a simple python script that can be distributed upon request. Supplementary Fig.~4 is based on ballistic trajectory calculations by one of the co-authors\cite{ElMellah2017MNRAS.467.2585E} solving the equation of motion using a classical fourth order Runge-Kutta scheme. The code is available for collaboration with I.E.M.\ upon reasonable request.
\end{addendum}


 \end{methods}

\afterpage\clearpage
\newpage 


%
%
%
%

\end{document}